\documentclass[1p]{elsarticle} 

\usepackage{subcaption} 
\usepackage{graphicx} 
\usepackage{circuitikz} 
\tikzset{>=latex} 
\usepackage{amsmath} 
\usepackage{xcolor}
\usepackage{xurl}
\usepackage{hyperref}

\newcommand{\simm}{\raise.17ex\hbox{$\scriptstyle\sim$}}
\newcommand{\bmat}{\begin{bmatrix}}
\newcommand{\emat}{\end{bmatrix}}
\newcommand{\setof}[1]{\left \{ #1 \right \}}
\newcommand{\midbar}{\ \middle | \ }

\newcommand{\ManualJToThermostat}{2.26}
\newcommand{\ManualJToThermostatLow}{0.87}
\newcommand{\ManualJToThermostatHigh}{5.21}

\newcommand{\RookStackToThermostat}{1.61}
\newcommand{\RookStackToThermostatLow}{0.80}
\newcommand{\RookStackToThermostatHigh}{2.75}

\newcommand{\HeuristicToThermostat}{4.39}
\newcommand{\HeuristicToThermostatLow}{1.81}
\newcommand{\HeuristicToThermostatHigh}{12.89}

\newcommand{\ExistingCapToThermostat}{2.75}
\newcommand{\ExistingCapToThermostatLow}{1.34}
\newcommand{\ExistingCapToThermostatHigh}{6.52}

\newcommand{\ManualJToBill}{2.35}
\newcommand{\ManualJToBillLow}{1.02}
\newcommand{\ManualJToBillHigh}{5.28}

\newcommand{\RookStackToBill}{1.70}
\newcommand{\RookStackToBillLow}{0.96}
\newcommand{\RookStackToBillHigh}{3.48}

\newcommand{\HeuristicToBill}{4.56}
\newcommand{\HeuristicToBillLow}{1.77}
\newcommand{\HeuristicToBillHigh}{10.19}

\newcommand{\ExistingCapToBill}{2.96}
\newcommand{\ExistingCapToBillLow}{1.36}
\newcommand{\ExistingCapToBillHigh}{7.76}

\newcommand{\PercentOfInterpolating}{72.9}

\begin{document}
\begin{frontmatter}
\title{Data-driven estimation of design heating loads for HVAC equipment sizing}

\author[me]{Alex H. Lee}
\author[trane]{Elias N. Pergantis}
\author[me,ece]{Kevin J. Kircher}

\address[me]{School of Mechanical Engineering, Purdue University, West Lafayette, IN 47907, USA}
\address[trane]{Trane Technologies, Residential R\&D Group, Tyler, TX 75707, USA}
\address[ece]{Elmore Family School of Electrical and Computer Engineering, Purdue University, West Lafayette, IN 47907, USA}

\begin{abstract}
Oversized heating and cooling equipment can unnecessarily increase up-front costs, energy costs, pollutant emissions, and strain on electrical infrastructure. This paper develops two data-driven methods for estimating heating loads at design conditions to improve equipment sizing. One method uses smart thermostat data; the other uses utility bills. We test the methods on a dataset that we gathered from 74 detached single-family houses in five United States climate zones. The dataset includes smart thermostat time-series data, monthly utility bills, weather data, existing equipment specifications, and Manual J design load calculations (the United States industry standard) purchased from practitioners. The two methods have strong goodness-of-fit statistics individually and show fair agreement with each other. On average over the 74 houses, existing heating equipment is \ExistingCapToThermostat\ (2.5th to 97.5th empirical percentile: \ExistingCapToThermostatLow--\ExistingCapToThermostatHigh) times larger than the thermostat-method estimate and \ExistingCapToBill \ (\ExistingCapToBillLow--\ExistingCapToBillHigh) times larger than the bill-method estimate. As implemented by practitioners, the Manual J estimate is \ManualJToThermostat\ (\ManualJToThermostatLow--\ManualJToThermostatHigh) times larger on average than the thermostat-method estimate and \ManualJToBill\ (\ManualJToBillLow--\ManualJToBillHigh) times larger than the bill-method estimate. We discuss prospects for implementing the data-driven methods at scale and for incorporating data-driven sizing into industry standards.
\end{abstract}

\end{frontmatter}

\section{Electrification and heat pump sizing}

Heating, ventilation, and air conditioning (HVAC) in residential and commercial buildings cause an estimated 15\% of global greenhouse gas (GHG) emissions and energy costs on the order of \$1 trillion\footnote{All monetary figures in this paper are inflation-adjusted to 2025 United States dollars.} per year worldwide \cite{khabbazi2025lessons}. Electrification---replacing fossil-fueled equipment with electric versions---is a leading approach in the academic literature \cite{Pistochini2022, Deetjen2021us}, non-governmental organizations \cite{ETC2025zero}, and governments \cite{DOE2024DPAHeatPump} to addressing GHG emissions from HVAC. Heat pumps (efficient electric heating and cooling machines) therefore play a central role in many climate action plans \cite{IEA2025heatpumps, ACEEE2024}.

This paper aims to reduce GHG emissions and energy costs by developing and testing data-driven methods to more appropriately size HVAC equipment in general and heat pumps in particular. This paper focuses primarily on heating in residential buildings, but the methods developed here could be extended to cooling and to small commercial buildings. This paper focuses on estimating design loads using data from the heating equipment that heat pumps typically replace in electrification projects, such as fossil-fueled furnaces and boilers.

\subsection{Issues with oversized heat pumps} \label{oversized_introduction}

Heat pumps have the potential to reduce GHG emissions significantly \cite{Dong2023, Gaur2021}, especially when replacing older, less efficient equipment and in locations with less carbon-intensive electricity \cite{Walker2022}. For three reasons, however, it is important that a heat pump is sized appropriately for the thermal loads it will serve. First, oversized heat pumps typically have higher upfront costs, as larger equipment tends to be more expensive to procure and install. Second, oversized heat pumps typically operate less efficiently, leading to higher GHG emissions and energy bills \cite{Bagarella2016, Carrier2025}. A heat pump's seasonal coefficient of performance (COP, the ratio of output thermal energy to input electrical energy) is typically highest when its capacity at design conditions is within $\pm$20\% of the building's design thermal loads \cite{Dongellini2017}. Third, an oversized heat pump can increase a building's peak electricity demand by up to 13\% \cite{Cummings2014, NREL2002}. Electricity demand peaks can strain electrical infrastructure within the building or on the power grid, potentially forcing costly replacement of circuit breaker panels, utility service drops, power lines, or transformers \cite{pergantis2025protecting, priyadarshan2025distribution}.

Estimates vary as to the frequency and severity of HVAC equipment oversizing. For example, Rhodes et al. estimated that 31\% of 4,971 air conditioners studied in Austin, Texas were oversized \cite{Rhodes2011}. Lucas et al. estimated that nearly half of 60 air conditioners studied in the Pacific Northwest were oversized beyond the typical 25\% safety margin \cite{Lucas1992}. Brudermueller et al. \cite{Brudermueller2025} estimated that 10\% of the heat pumps installed in central Europe were oversized by a factor of two or more.

\subsection{Existing methods for equipment sizing}

Today, HVAC equipment sizing is often done by contractors, who may prefer to oversize equipment to avoid complaints, accommodate future building expansions, or allow faster response to temperature setpoint changes \cite{Vieira1996}. Some studies show that 30\% to more than half of contractors do not perform a formal load calculation when sizing equipment \cite{Vieira1996,Sullivan2025}. Many contractors size new equipment using a rule of thumb based on a floor area, such as 350 to 700 ft$^2$ per ton (9.25 to 18.5 m$^2$/kW) of rated cooling capacity \cite{Vieira1996, Burdick2011}, or simply replace the existing unit with a new one with the same rated capacity \cite{ACCA2018} (referred to here as like-for-like replacement). These practices often lead to oversizing, and therefore to the cost and performance issues outlined in Section \ref{oversized_introduction}.

More rigorous HVAC equipment sizing typically uses physics-based estimates of the building's heating and cooling loads under design indoor and outdoor conditions. For example, Manual J is the American National Standards Institute standard for estimating design loads for single-family detached houses, small multi-unit residences, condominiums, townhouses, and manufactured housing \cite{ACCA_ManualJ}. The Manual J method begins by specifying the building geometry, material properties, outdoor air infiltration rates, and other themophysical parameters. Manual J then uses these parameters to calculate steady-state heat transfer through the building envelope under design boundary conditions.

While Manual J and similar energy balance methods generally lead to better sizing decisions than rules of thumb, they have several potential drawbacks. First, it can be difficult to find a properly trained individual to perform energy balance calculations. Many HVAC contractors do not offer Manual J estimates, instead relying on rules of thumb or like-for-like replacement \cite{Peavy2025, Holladay2024}. Second, energy balance calculations can be expensive. One source states that Manual J estimates typically cost around \$100 to \$300 for smaller houses and up to \$1,000 for larger and more complex houses \cite{ResCheck2024}, while another source mentions \$1,500 to \$2,500 \cite{Lunsford2025}. Third, it can be difficult to specify accurate input data to energy balance calculations due to unknown construction materials, undocumented renovations to the building envelope, limited access to crawl spaces, inaccurate duct schematics, or other challenges \cite{DOE2018}. This makes energy balance calculations vulnerable to human errors or negligence, as inaccurate inputs can lead to inaccurate outputs. Fourth, despite its susceptibility to uncertainty, Manual J produces only point estimates of design loads. A more robust method would return uncertainty information alongside point estimates.

The need for more accessible, affordable, and robust HVAC equipment sizing methods is widely recognized \cite{DOE2018}. In recent years, several companies \cite{Farrell2024, Rookstack} and nonprofits \cite{EfficiencyMaine} have developed software tools for estimating design loads. Two of these tools \cite{Farrell2024, EfficiencyMaine} use fuel consumption data from utility bills and one \cite{Rookstack} uses a housing characteristics survey and public datasets. These tools provide accessible means for estimating design thermal loads, but their accuracy has not been evaluated by a neutral third party. While this paper focuses on developing data-driven methods for design load estimation, it also evaluates one commercial software tool \cite{Rookstack}.

This paper develops and tests two data-driven methods that advance the state of the art of HVAC equipment sizing. Both methods require less human-specified input data---at minimum, the make and model of the existing HVAC equipment, the zip code, and either utility bills or historical data from a smart thermostat---and so are less susceptible to data acquisition errors. Both methods use data from the actual building envelope and thermal distribution system. These data capture factors that affect real-world performance but are difficult to specify precisely for other sizing methods, such as rates of outdoor air infiltration or supply air leakage from distribution ducts in unconditioned space. Providing accurate input data to the methods developed here requires no expertise in building science, making them potentially more accessible and affordable than energy balance methods. Both data-driven methods also have the potential to be fully automated, enabling anyone with access to the input data to run the methods. The methods developed here use few, simple equations, making them interpretable and facilitating error diagnosis and uncertainty quantification.

\subsection{Contributions and potential impact}

This paper makes four main contributions to the research literature on design-load estimation for HVAC equipment sizing. First, this paper builds a new, open dataset from 74 detached single-family houses in five United States climate zones. For each house, this dataset includes housing characteristics, Manual J design load estimates, existing HVAC equipment specifications, utility bills, time-series data from a smart thermostat, and weather data. This is the largest and richest open dataset of this type of which the authors are aware.

Second, this paper develops and tests two data-driven methods for estimating design thermal loads. The first, referred to as the thermostat method, uses the existing HVAC equipment capacity and time-series data from a smart thermostat. The second, referred to as the bill method, uses the existing HVAC equipment efficiency and monthly utility bills. While versions of both methods can be found in the research literature or professional practice, this paper makes several methodological refinements that can improve accuracy and robustness. In the test data, the thermostat and bill methods have strong goodness-of-fit statistics individually and show fair agreement with each other.

Third, this paper compares the thermostat and bill methods to today's sizing methods, including Manual J, a commercial software tool, a rule of thumb based on floor area, and like-for-like replacement. We find that on average over the 74 houses, existing heating equipment is \ExistingCapToThermostat\ times larger than the thermostat-method estimate and \ExistingCapToBill\ times larger than the bill-method estimate. Manual J, as implemented today by practitioners, overestimates design heating loads on average by factors of \ManualJToThermostat\ relative to the thermostat method and \ManualJToBill\ relative to the utility bill method. The commercial software tool is generally about midway between the data-driven methods and Manual J. The floor-area rule of thumb produces much higher design load estimates than all other methods.

Fourth, this paper reports lessons learned from gathering the dataset and from implementing a portfolio of sizing methods on the 74 houses. Key lessons include: (a) Finding enthusiastic participants for an HVAC study is a real challenge. To build a full dataset for 74 houses, we reached out to more than 5,000 people. (b) Manual J estimates are difficult and expensive to obtain. Of the 110 residential energy specialists we contacted in 18 United States cities, over 80\% declined to provide a quote for Manual J services. For the specialists who did provide quotes, the cost of a Manual J estimate ranged from \$300 to \$2,500. Of the 16 specialists we paid for Manual J calculations, 14 visited the physical house. Only one did a blower-door test\footnote{The Manual J calculation that included a blower-door test cost \$1,200 (after a 25\% academic discount on the original quote of \$1,600). This was the most expensive Manual J calculation we purchased. It was significantly closer than the average Manual J calculation to the data-driven estimates: 1.50 and 1.65 times higher than the thermostat- and bill-method estimates for the house, respectively.} to estimate the outdoor air infiltration rate, an influential and hard-to-specify input to Manual J calculations. (c) Of the two data-driven methods developed here, the bill method is likely better suited to implementation at scale. Almost every household has utility bills, while relatively few have smart thermostats that can log and export data.

The potential impact of these contributions is to build confidence among decision-makers in business and government that data-driven methods can provide reliable estimates of design thermal loads. This could enable decision-makers to update HVAC equipment sizing standards to allow or require data-driven methods. Updated standards could shift industry practices and reduce rates of oversizing. Reduced rates of oversizing could reduce up-front equipment costs and improve energy efficiency, thereby reducing utility bills, pollutant emissions, and strain on electrical infrastructure.

The rest of this paper is structured as follows. Section \ref{related_works} summarizes related research. Section \ref{methodology} details the methods used for data collection and data-driven design load estimation. Section \ref{results} presents the main results. Section \ref{discussion} discusses implementation practicalities and lessons learned from data collection. Section \ref{conclusion} concludes the paper.

\section{Related work}
\label{related_works}

This section contextualizes the work in this paper relative to past research on HVAC equipment sizing and on data-driven thermal modeling of buildings. Section \ref{designLoadReview} reviews existing methods for estimating design thermal loads. Section \ref{lit_review_data_driven_model} surveys physics-based, data-driven, and hybrid methods for modeling building temperature dynamics. Section \ref{thermostatStudyReview} focuses in more detail on closely related studies that use field data from residential thermostats.

\subsection{Existing design-load estimation methods}
\label{designLoadReview}

Existing methods for estimating thermal loads at design conditions fall into three broad categories \cite{Gang2015}. First, practitioners often use rules of thumb based on climate zone and floor area. These rules of thumb often lead to oversizing \cite{Vieira1996} and have changed little over time, even as building insulation has improved and energy codes have increased in stringency \cite{Burdick2011}.

The second category includes linear steady-state heat transfer methods such as Manual J. These methods use geometry and material properties to estimate heat transfer through the building envelope from conduction and outdoor air infiltration, as well as internal gains from body heat, electrical loads, and the sun \cite{acca2016manualj}. While Manual J can have high accuracy given accurate input data \cite{Lucas1992, Munk2014, Proctor2006}, in practice, inaccurate input data can degrade estimation accuracy significantly \cite{Burdick2011, DOE2018}.

The third category includes nonlinear transient simulation software such as EnergyPlus, DOE-2, TRNSYS, and DeST \cite{Pan2023, Gasparella2013}. These simulators generally require more extensive data on geometry and material properties than linear steady-state methods. Transient simulators produce time-series estimates of thermal loads corresponding to input weather data, such as a typical meteorological year or a design-day profile \cite{Gang2015, Hong1999}. Time-series load estimates enable equipment selection based on design loads or life-cycle performance assessment. As with linear steady-state methods, nonlinear transient simulation methods generally require expert domain knowledge to run well, and their accuracy depends on input data quality \cite{Megri2007}.

\subsection{Data-driven building thermal modeling} \label{lit_review_data_driven_model}

Data-driven models can be an alternative or complement to so-called white-box models, such as linear steady-state heat transfer models and nonlinear transient simulators, which are based only on physics. Data-driven models use measured data from the building, so can more accurately capture as-built performance \cite{ASHRAE2025Ch19}. Data-driven models can be categorized as black-box or gray-box \cite{ASHRAE2025Ch19}. Black-box models are purely data-driven, while gray-box models fuse data and physics. The review article \cite{AbdelJaberDirks2024} argues that while black-box models can accurately predict building loads and reveal energy efficiency opportunities, they often lack interpretability. Another review highlights challenges in applying black-box models to residential buildings, which may lack high-quality sensor data or show highly variable occupant behavior \cite{AmasyaliElGohary2018}.

A popular gray-box technique for building thermal modeling uses a resistance-capacitance (RC) thermal circuit structure \cite{Ogunsola2012, Li2021, Boodi2022}. RC models represent thermal dynamics by analogy to electrical circuits, with heat playing the role of charge and temperature playing the role of voltage. Capacitors represent thermal zones (such as a room or floor) or building components (such as a wall, window, or roof), while resistors represent things that impede heat transfer (such as wall insulation). Gray-box RC methods use data-driven optimization techniques to search for parameter values that minimize errors between model predictions and measured data \cite{Ogunsola2012}. 

The gray-box tradition most relevant to this paper models a building's thermal load as linear in the indoor-outdoor temperature difference. This tradition traces back at least to the Princeton Scorekeeping Method (PRISM) \cite{fels1986prism}, which researchers developed in the late 1970s \cite{schrader1978two} to provide weather-normalized energy scorecards before and after energy efficiency retrofits. PRISM linearly regresses fuel use on heating degree-days, drawing inspiration from a steady-state energy-balance model similar to the dynamic RC model used in this paper. Other methods in this tradition are variously referred to as load lines, building performance lines, energy signatures, or degree-day analyses \cite{day2003improved}. Building on that long tradition, this paper applies Bayesian bootstrapping techniques to PRISM-style load line methods to improve estimation of confidence intervals (CIs), extends the linear regression approach to high-resolution time-series data from smart thermostats, applies the refined methods to the equipment sizing problem (rather than, for example, to more general load forecasting or to energy efficiency assessment), and tests them on a new, rich dataset that we gathered.

\subsection{Residential thermostat field studies} \label{thermostatStudyReview}

The thermostat-based data-driven design load estimation method in this paper builds on prior work from seven recent studies that used smart thermostat data \cite{Huchuk2021, Li2024, Vallianos2024, Hossain2021, Baasch2019, Doma2023, Zeifman2020} . The first six studies used data from a popular thermostat manufacturer's Donate Your Data (DYD) program \cite{ecobeeDYD}. The opt-in DYD program collects field data including city, state or providence, temperature settings and overrides, occupancy schedules, indoor temperatures, and HVAC equipment runtimes. The DYD dataset does not include HVAC equipment capacities or efficiencies. The studies that used the DYD dataset typically fit for lumped RC parameters, such as the product of the resistance term $R$ ($^\circ$C/kW) and the unknown capacity of the HVAC equipment $K$ (kW, often assumed to be constant). The seventh study \cite{Zeifman2020} used a utility program dataset that includes data from three smart thermostat vendors, as well as HVAC equipment capacities and home energy assessment results.

Four of the studies \cite{Huchuk2021, Li2024, Vallianos2024, Hossain2021} focused on developing models for advanced control of HVAC equipment. Huchuk et al. \cite{Huchuk2021} used data from 1,000 thermostats in the United States; Li et al. \cite{Li2024} from 164 thermostats in the California, Texas, and New York; Vallianos et. al \cite{Vallianos2024} from 60,000 thermostats across North America; and Hossain et al. \cite{Hossain2021} from 8,884 thermostats in Canada. These studies developed a mix of black-box \cite{Huchuk2021, Li2024} and gray-box models \cite{Huchuk2021, Vallianos2024, Hossain2021}.

The other three studies \cite{Baasch2019, Doma2023, Zeifman2020} focused on identifying houses with high potential benefits from energy-efficiency retrofits. Baasch et al. \cite{Baasch2019} compared three methods for fitting RC models for 4,000 houses in Ontario and New York. Out of the three methods, the balance-point method is most similar to the thermostat method we develop in this paper. The balance-point method linearly regresses the daily-average equipment runtime, a stand-in for the heat supply that is not computable from the DYD dataset, against outdoor temperature. Baasch et al. noted that the linear fits, while generally statistically significant, had high standard errors and weak coefficients of determination. In this paper, we add a data pre-processing step that selects overnight data, rather than using daily averages as in \cite{Baasch2019}. This step mitigates uncertainty from internal gains due to body heat, electrical loads, and sunlight, improving model fits.

Doma et al. \cite{Doma2023} applied the decay-curve and energy-balance methods, which resemble the balance-point method in \cite{Baasch2019}, to 60,000 houses across North America. Doma et al. fit the product $RC$ (the time constant associated with the indoor temperature dynamics) and proposed using it to target retrofits.

Zeifman et al. \cite{Zeifman2020} used a utility program dataset, including smart thermostat data from 87 houses in Massachusetts, to fit a second-order RC model for each house. Before fitting the RC parameters, Zeifman et al. integrated time-series data over long periods of time, reducing the time-series estimation problem to a static one and mitigating the effects of thermal response delays and disturbances from internal heat gains. This paper uses a similar time-integration approach, although (as discussed above) we restrict our attention to nights, when solar effects and variability in occupant behavior are minimal. Zeifman et al. compared their estimates of the overall thermal resistance between the indoor and outdoor air to physics-based estimates from home energy assessments. They found that their method had 88\% (70 out of 80) accuracy in classifying houses with poor vs. adequate insulation.

\section{Methods}\label{methodology}

This section describes data collection (Section \ref{method_dataCollection}), as well as the thermal circuit model (Section \ref{buildingModel}) that underlies both the thermostat method (Section \ref{method_thermostat}) and the utility bill method (Section \ref{bill_method}) for estimating design thermal loads.

\subsection{Data collection}
\label{method_dataCollection}

We aimed to collect a range of data from participants, including historical thermostat records, heating fuel use, HVAC equipment specifications, and responses to a survey on housing characteristics. To gauge willingness to participate and suitability of heating equipment, we sent a short survey to  university email lists, social media connections, and online building energy forums. We estimate the total number of recipients to this outreach at about 5,000, from which we received 316 responses. We screened the 316 respondents for suitability, selecting 125 who were willing to share all the relevant data and allow an on-site home energy evaluation; heated with natural gas, propane, or electric resistance; did not use secondary heating equipment; and either had a smart thermostat or were willing to have one installed.

\begin{figure}
    \centering
    \includegraphics[width=0.5\textwidth]{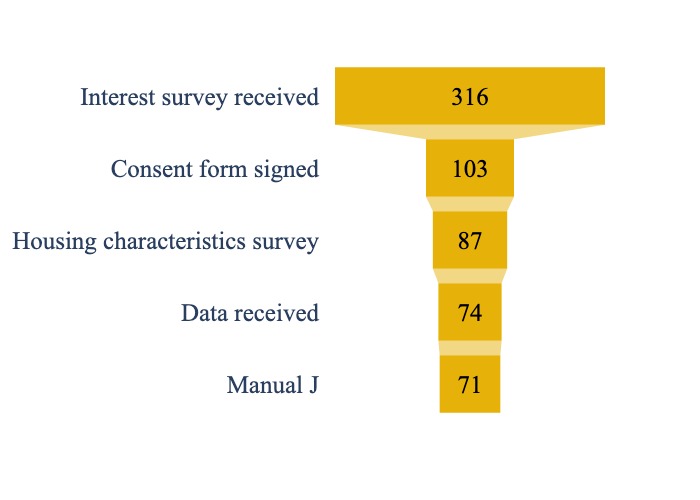}
    \caption{From an outreach of more than 5,000 people, 316 responded with interest. Of the 316,  we selected 103 to join the study and 71 completed the entire process.}
    \label{fig:participant_progress}
\end{figure}

We offered each of the 125 selected participants \$10 for completing the housing characteristics survey, \$10 for uploading utility bills and a photo of the heating equipment nameplate, \$10 for uploading thermostat time series, and \$50 for allowing an on-site visit for a Manual J calculation. We asked the participants to complete each step of the study sequentially, only asking for a new task if they completed the previous one. Figure \ref{fig:participant_progress} shows the order and the number of participants who completed each task. We tried to schedule as many in-person Manual J calculations as possible, but encountered many difficulties: Unresponsive participants, lack of nearby practitioners willing to perform a Manual J calculation, and quotes well beyond our budget. The most we paid for a single Manual J calculation was \$1,200. We received quotes as high as \$2,500.

Participant recruitment ran for about four months. Interaction with participants, such as acquiring thermostat data and equipment nameplate photos, went on for four more months. Scheduling in-person Manual J calculations took two to three months on average, including finding a practitioner in the area to perform the calculation and helping the participant schedule the visit.

In the end, we received utility bill data, thermostat data, and equipment nameplate photo from 74 participants and Manual J calculations from 71 participants. We verified completeness and integrity of the dataset from 70 participants, as some participants had missing or corrupted data (such as no Manual J calculations or a smart thermostat time series with unacceptably large gaps) that we could not resolve.

\subsection{Thermal circuit model}
\label{buildingModel}

\begin{figure}
\centering
\begin{circuitikz}[scale=1.1, american currents] 
\ctikzset{bipoles/length=1.05cm} 
\pgfmathsetmacro{\w}{2};
\pgfmathsetmacro{\h}{1};

\node[above] at (2*\w,2*\h) {$T_\text{out}$};
\draw (2*\w,2*\h) to[battery1,*-] (2*\w,0) -- (0,0);

\node[above] at (\w,2*\h) {$T_\text{in}$};
\draw (0,2*\h) to (\w,2*\h) to[R,R=$R$,*-] (2*\w,2*\h);

\draw (0,0) to[I,n=Qa] (0, 2*\h);
\node[right] at (Qa.s) {$\dot Q_h + \dot Q_e$};

\draw (\w,0) to[C,n=Ca] (\w,2*\h);
\node[right] at (Ca.s) {$C$};
\draw (\w,0) node[ground] {} to (\w,0);

\end{circuitikz}
\caption{The 1R1C thermal circuit model underlying the thermostat and bill methods. In the analogy to electrical circuits, temperature plays the role of voltage and thermal power plays the role of current.}
\label{1R1CFig}
\end{figure}
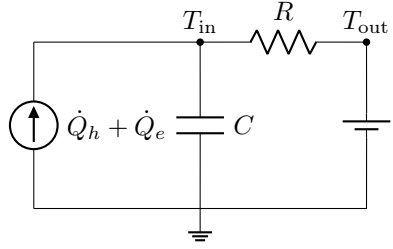

The first-order thermal circuit model structure in Figure \ref{1R1CFig} underlies both the thermostat method and the energy bill method. In this model, thermal power plays the role of current and temperature plays the role of voltage. The indoor air has temperature $T_\text{in}$ ($^\circ$C) and capacitance $C$ (kWh/$^\circ$C). The thermal resistance $R$ ($^\circ$C/kW) impedes heat transfer between the indoor air and the outdoor air, which has temperature $T_\text{out}$ ($^\circ$C). Heating equipment supplies thermal power $\dot Q_h$ (kW) to the indoor air. Exogenous heat sources---such as body heat, lights, appliances, and the sun---supply thermal power $\dot Q_e$ (kW) to the indoor air. The first-order linear ordinary differential equation
\begin{equation}
C \frac{\text d T_\text{in}(t)}{\text d t} = \frac{T_\text{out}(t) - T_\text{in}(t)}{R} + \dot Q_h(t) + \dot Q_e(t)  \label{eq:ct1R1C}
\end{equation}
governs the indoor air temperature dynamics.

\subsection{Thermostat method}\label{method_thermostat}

The thermostat method uses historical data from a smart thermostat to train the parameters of the thermal circuit model, Equation \eqref{eq:ct1R1C}. Smart thermostats typically keep a five-minutely record of various temperature, humidity, and occupancy measurements. We used measurements of the indoor temperature $T_\text{in}$, outdoor temperature $T_\text{out}$, and how long the heating equipment ran in each five-minute measurement interval. Assuming the heating equipment supplies constant thermal power whenever it runs, $\dot Q_h(k) = \psi(k) \dot Q_{\text{cap}}$, where $\psi$ is the (dimensionless) run-time fraction, $\dot Q_{\text{cap}}$ (kW) is the equipment's heating capacity, and the integer $k$ indexes measurement windows. Therefore, only the parameters $C$, $R$, and $\dot Q_e$ in Equation \eqref{eq:ct1R1C} are unknown. The thermostat method time-averages Equation \eqref{eq:ct1R1C} to eliminate $C$, fits $R$ and $\dot Q_e$, and evaluates the trained thermal circuit model at the design indoor and outdoor temperatures to estimate the design heating load.

\subsubsection{Time-averaging to eliminate $C$} \label{thermostat_time_average}

Time-averaging Equation \eqref{eq:ct1R1C} over any interval $[t_0,t_\text{end}]$ with $T_\text{in}(t_0) = T_\text{in}(t_\text{end})$ eliminates the thermal capacitance $C$. To see this, we integrate both sides of Equation \eqref{eq:ct1R1C} with respect to time:
\begin{equation}
\begin{aligned}
&C \int_{t_{\text{0}}}^{t_\text{end}}\frac{\text d T_\text{in}(t)}{\text d t} \text d t \\
& = \int_{t_{\text{0}}}^{t_\text{end}} \left[\frac{T_\text{out}(t) - T_\text{in}(t)}{R} + \dot Q_h(t) + \dot Q_e(t)\right] \text d t .\label{1R1C_integral}
\end{aligned}
\end{equation}
By the fundamental theorem of calculus, the left-hand side reduces to $C \left( T_\text{in}(t_\text{end}) - T_\text{in}(t_{\text{0}}) \right)$, which vanishes since $T_\text{in}(t_0) = T_\text{in}(t_\text{end})$. Therefore,
\begin{equation}
q_h = 
\frac{\Delta T}{R}  - q_e , \label{1R1C_simplified}
\end{equation}
where
\begin{equation}
\begin{aligned}
q_h &= \frac{1}{t_\text{end} - t_0} \int_{t_0}^{t_\text{end}} \dot Q_h(t) \text d t \\
\Delta T &= \frac{1}{t_\text{end} - t_0} \int_{t_0}^{t_\text{end}} \left( T_\text{in}(t) - T_\text{out}(t) \right) \text d t \\
q_e &= \frac{1}{t_\text{end} - t_0} \int_{t_0}^{t_\text{end}} \dot Q_e(t) \text d t .
\end{aligned}
\end{equation}

\subsubsection{Data pre-processing} \label{thermostat_preprocess}

After time-averaging, the task of data pre-processing reduces mainly to identifying periods with roughly equal initial and final indoor temperatures. We further restrict the data to overnight periods, which simplifies the exogenous thermal power $\dot Q_e$ by eliminating solar effects and reducing variability in occupant behavior. Night data are also more relevant to design heating loads, which typically occur in the hours before sunrise.

\begin{figure}
\centering
\includegraphics[width=0.5\textwidth]{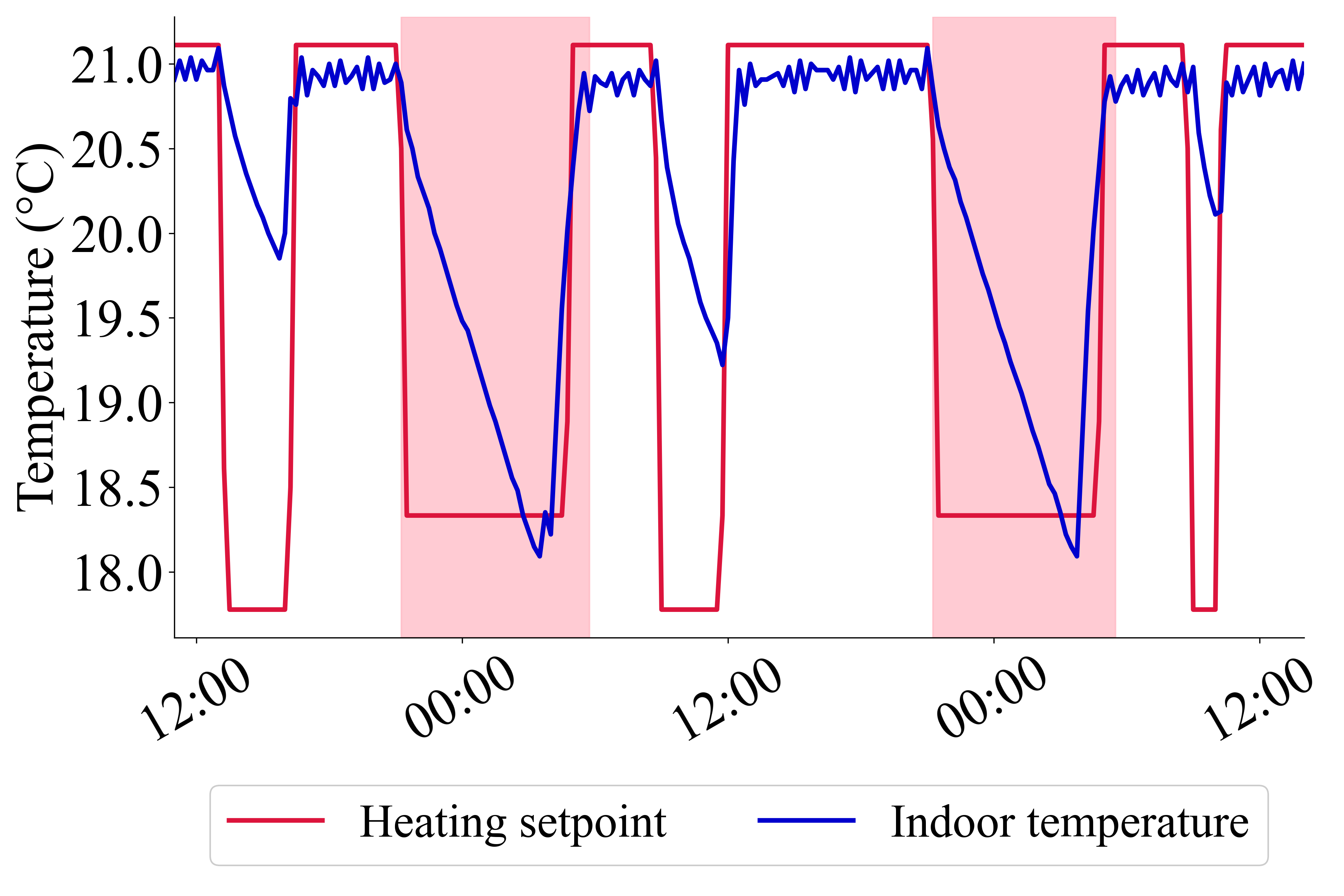}
\caption{Data pre-processing for night setback schedules. Each pink-shaded period is one night, within which data are time-averaged before model fitting.}
\label{dataSelection}
\end{figure}

In our study, 82\% of participants reduced their heating temperature setpoints overnight. To select overnight periods with roughly equal initial and final indoor temperatures for these participants, we start at the beginning of the night, just prior to the setpoint reduction, and end the window just after the indoor temperature returns to the day-time setpoint. Figure \ref{dataSelection} shows two selected intervals (pink shaded areas) for one participant. Within each selected interval, we time-average the temperature and power measurements into a single observation of $(\Delta T, q_h)$.

\subsubsection{Model fitting} \label{thermostat_model_fitting}

Given observations ($\Delta T_i,  q_{h,i})$ for nights $i = 1$, \dots, $n$, the unknowns $1/R$ and $q_e$ in Equation \ref{1R1C_simplified} can be found using linear regression with the model
\begin{equation}
    \begin{aligned}
    y= X\boldsymbol{\beta} + \varepsilon ,
    \label{Thermostat_regression}
\end{aligned}
\end{equation}
where 
\begin{equation}
    \nonumber
    \begin{aligned}
    y =
    \begin{bmatrix}
        q_{h, 1} \\
        \vdots \\
        q_{h, n}
    \end{bmatrix}, \ X=
    \begin{bmatrix}
        \Delta T_1   & -1\\
        \vdots  & \vdots \\
        \Delta T_n & -1
    \end{bmatrix}, \ \beta =
    \begin{bmatrix}
        1/R \\
        q_e \\
    \end{bmatrix} ,
\end{aligned}
\end{equation}
and $\varepsilon$ is an $n$-dimensional vector of prediction errors. The least-squares estimate $\hat \beta$ of $\beta$ minimizes the Euclidean norm of the error $y - X \beta$. To ensure that the fitted parameters are consistent with their underlying physical interpretations, we constrain $q_e$ and $R$ to be nonnegative.

\subsubsection{Design load estimation}
\label{thermostat_design_load}

The design heating load estimate is
\begin{equation}
    \begin{aligned}
    \hat q_{h,\text{des}} = \hat \beta_1 \Delta T_\text{des} - \hat \beta_2 ,
\end{aligned}\label{design_point_estimate}
\end{equation}
where $\Delta T_\text{des}$ ($^\circ$C) is the design indoor-outdoor temperature difference, $\hat \beta_1$ (kW/$^\circ$C) is the estimate of $1/R$, and $\hat \beta_2$ (kW) is the estimate of $q_e$. We view $\hat q_{h,\text{des}}$ (kW) as the estimated thermal power required to keep the indoor temperature constant at its design value under the design outdoor temperature.

\subsubsection{Uncertainty quantification} \label{Thermostat_design_CI}

\begin{figure}
    \centering
    \includegraphics[width=0.5\textwidth]{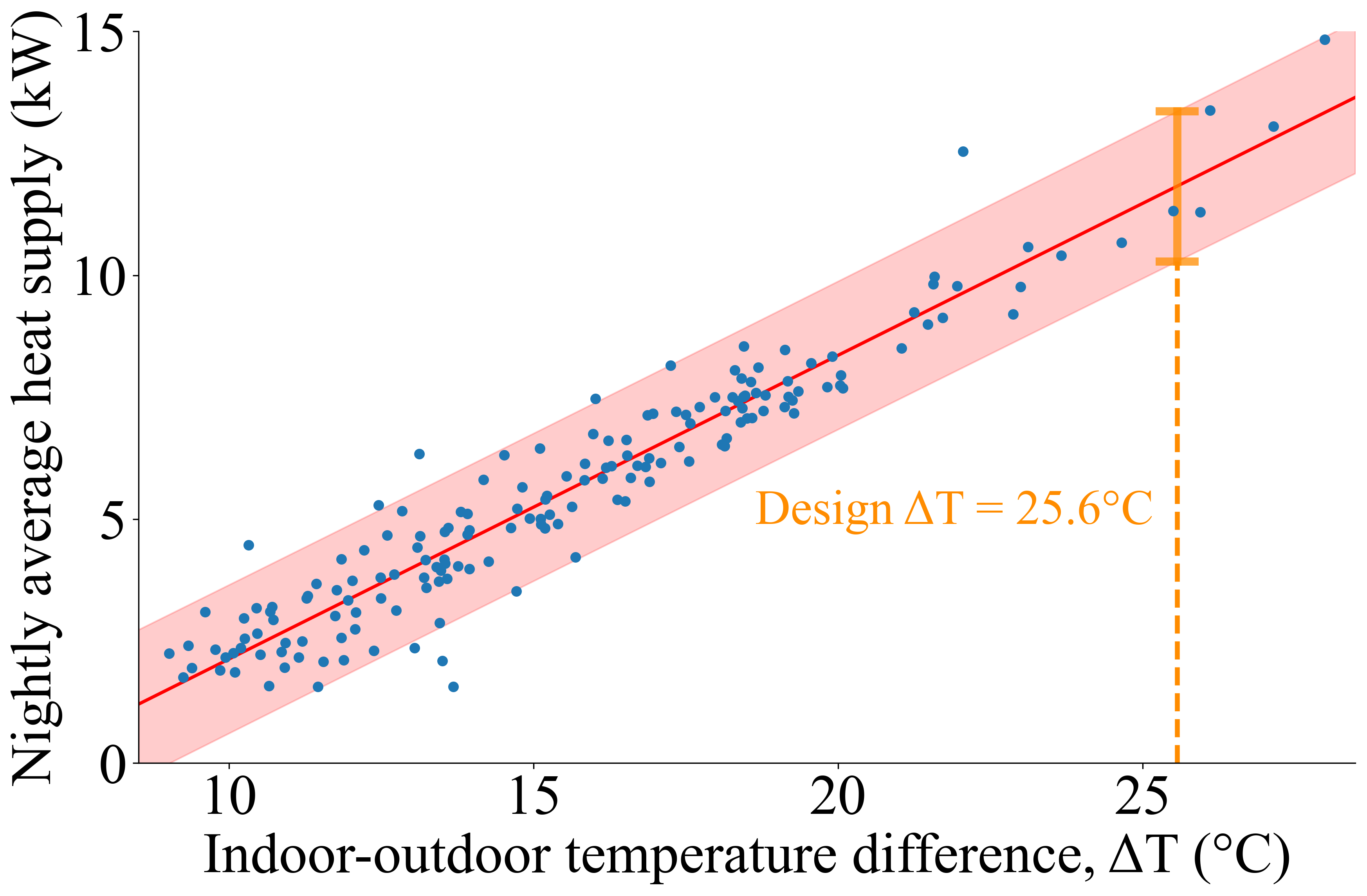}
    \caption{Thermostat method example. Red band: 95\% CI of heating load at each indoor-outdoor temperature difference. Orange interval: 95\% CI at design condition.}
    \label{fig:Thermostat_CI}
\end{figure}

Eq. \eqref{design_point_estimate} provides a point estimate of the design heating load, but accompanying point estimates with uncertainty information can improve HVAC sizing decisions. If the error $\varepsilon$ in Equation \eqref{Thermostat_regression} is Gaussian, then the CI for the design load estimate is
\begin{equation}
    \begin{aligned}
    \hat q_{h,\text{des}} \pm t\left(1- \alpha/2, n-2\right) \hat \sigma_\text{des} ,
\end{aligned}\label{thermostat_CI}
\end{equation} 
where $t(1-\alpha/2, n-2)$ is the critical $t$-value from a Student's $t$-distribution with $n - 2$ degrees of freedom at significance level $\alpha$. The standard error for a new prediction at the design indoor-outdoor temperature difference $\Delta T_\text{des}$ is
\begin{equation}
    \begin{aligned}
    \hat \sigma_\text{des} = \hat \sigma\sqrt{  1 + \frac{1}{n} + 
    \frac{\Big(\Delta T_\text{des} - \overline{\Delta T}\Big)^2}
         {\displaystyle\sum_{i=1}^n \Big(\Delta T_{i} - \overline{\Delta T}\Big)^2} } ,
    \end{aligned}
    \label{pred_std_err}
\end{equation}
where $\hat \sigma$ is the standard error of the regression residuals and $\overline{\Delta T} = (\Delta T_1 + \dots + \Delta T_n)/n$ \cite{Kutner2004Chapter2_5}. Figure \ref{fig:Thermostat_CI} shows the 95\% CI (pink shaded area) around the fitted line (red) for one of the participant's data, as well as the CI for the design heating load (orange vertical interval).

\subsection{Utility bill method}
\label{bill_method}

The utility bill method works similarly to the thermostat method, but uses fuel consumption and the equipment efficiency to derive the heat supplied to the house, rather than thermostat run-time data and the equipment capacity. The dataset used here consists of monthly natural gas bills, but the method could also work with electricity bills for houses with electric resistance heating or heat pumps for which accurate COP curves are available. We asked participants to upload at least one year of natural gas bills. Most participants did, but a few could only provide several months of data because they recently moved.

The bill method uses the same 1R1C model as the thermostat method. Given a billing interval $[t_0,t_\text{end}]$, we assume that occupants turn on the heating equipment only when the outdoor temperature is below a threshold $\theta$ ($^\circ$C), i.e., when 
\begin{equation}
t \in \mathcal T = \setof{t \in [t_0, t_\text{end}] \midbar T_\text{out}(t) \leq \theta } . 
\end{equation}
Integrating Equation \eqref{eq:ct1R1C} over $\mathcal T$ and assuming the left-hand side vanishes as in Section \ref{thermostat_time_average} gives
\begin{equation}
Q_h = \frac 1 R \int_{t \in \mathcal T} ( T_\text{in}(t) - T_\text{out}(t) ) \text d t  - Q_e , \label{bill_intermediate}
\end{equation}
where
\begin{equation}
Q_h = \int_{t \in \mathcal T} \dot Q_h(t) \text d t , \ Q_e = \int_{t \in \mathcal T} \dot Q_e(t) \text d t .\label{bill_heat_supplied} \\
\end{equation}
We model $Q_h$ (kWh), the heat supplied over the billing interval, as $Q_h = \eta E$, where $\eta$ is the (dimensionless) heating equipment efficiency and $E$ (kWh) is the energy content of the fuel used for space heating over the interval. We obtain $\eta$ from equipment specifications and assume it is time-invariant.

In designing the bill method, we do not assume that time-series observations of $T_\text{in}(t)$ are available. Evaluating the integral in Equation \eqref{bill_intermediate} therefore requires making a simplifying assumption about the indoor temperature. Here we choose $\theta$ to be a representative indoor temperature and assume $T_\text{in}(t) = \theta$ for all $t \in \mathcal T$. Under this assumption, 
\begin{equation}
\begin{aligned}
&\int_{t \in \mathcal T} ( T_\text{in}(t) - T_\text{out}(t) ) \text d t  = \int_{t \in \mathcal T} ( \theta - T_\text{out}(t) ) \text d t \\
= &\int_{t_0}^{t_\text{end}} ( \theta - T_\text{out}(t) ) f(t) \text d t \\
= &\int_{t_0}^{t_\text{end}} \max(0, \theta - T_\text{out}(t) ) \text d t .
\end{aligned}
\end{equation}
In the second line, the function $f(t)$ equals one if $T_\text{out}(t) \leq \theta$ (i.e., if $t \in \mathcal T$) and zero otherwise.

In summary, the bill method's final model is
\begin{equation}
q_h = \frac{\Delta T^+}{R} - q_e , \label{bill_final}
\end{equation}
where $q_h = \eta E /(t_\text{end} - t_0)$, $q_e = Q_e / (t_\text{end} - t_0)$, and
\begin{equation}
\Delta T^+ = \frac{1}{t_\text{end} - t_0} \int_{t_0}^{t_\text{end}} \max(0, \theta - T_\text{out}(t) ) \text d t .
\end{equation}
Eq. \eqref{bill_final} is a linear model similar to Equation \eqref{1R1C_simplified}. The fitting and design load estimation methods described in Sections \ref{thermostat_model_fitting}--\ref{thermostat_design_load} apply directly.

To generate the results in this paper, we set the parameter $\theta$ equal to the time-average of the heating temperature setpoint. For example, if a household keeps the heating temperature setpoint at 21 $^\circ$C from 6 AM to 10 PM and 18 $^\circ$C from 10 PM to 6 AM, then we set $\theta = 20$ $^\circ$C. When historical setpoint data are unavailable, a bill-method implementation could ask the user to upload the typical heating temperature setpoint when they upload the energy bills. Alternatively, the fitting procedure could treat $\theta$ as a tunable hyperparameter, sweep a range of candidate $\theta$ values, and select the value that gives the best fit in validation data.

\subsubsection{Data pre-processing}

To calculate $\Delta T^+$, we use location-specific outdoor temperature data from the Oikolab weather service \cite{oikolab}, which provides historical data at hourly time steps for any location based on nearby weather stations. Each billing period (typically one month) provides one observation of $(\Delta T^+, q_h)$.

Because the bill method's input data typically span a whole year, we filter out summer data and use only `heating months' for model fitting. As the definition of `heating months' varies over climates and occupants, we developed an automated categorization method that filters out months when the heating system is inactive. For each house, the method calculates a threshold $\tau^\star$ ($^\circ$C) such that when $\Delta T^+ \leq \tau^\star$, the heating equipment is inactive. We fit only to data from months with $\Delta T^+ > \tau^\star$.

\begin{figure}
    \centering
    \includegraphics[width=0.5\textwidth]{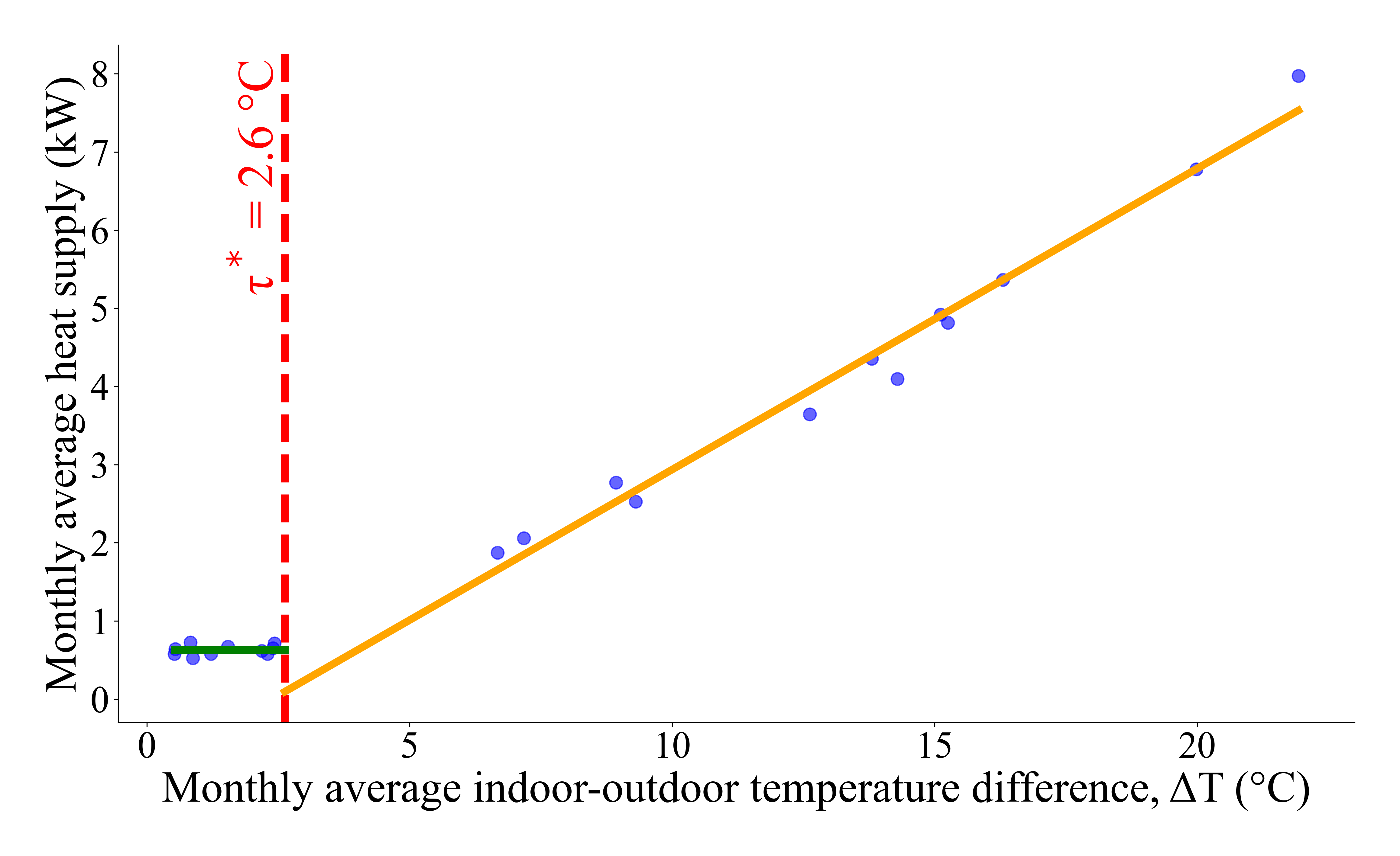}
    \caption{The threshold $\tau^\star$ minimizes the categorization error between heating and non-heating months. The model is trained only on months with mean indoor-outdoor temperature differences above $\tau^\star$.}
    \label{Bill_threshold}
\end{figure}

The heating-month selection algorithm begins by sorting months in increasing order of $\Delta T^+$. It then loops over months, indexed by $i$, testing the candidate threshold $\tau_i = \Delta T^+_i$. For each $i$, the algorithm bifurcates the months into a group with $\Delta T^+ \leq \tau_i$ and a group with $\Delta T^+ > \tau_i$. For each month with $\Delta T^+ \leq \tau_i$, we compute the root mean squared error (RMSE) between the month's heat supply and the mean heat supply over all months with $\Delta T^+ \leq \tau_i$. For each month with $\Delta T^+ > \tau_i$, we compute the RMSE between the month's heat supply and a linear fit of $q_h$ vs. $\Delta T^+$ over all months with $\Delta T^+ > \tau_i$. We select the threshold $\tau^\star$ that minimizes the sum of the two RMSEs over all the candidate $\tau_i$ values.

The main idea behind the heating month selection algorithm is that for non-heating months, fuel use corresponds to activities other than space heating, such as water heating, cooking, or drying clothes. We assume that fuel use for these non-space-heating activities is constant over months and independent of the outdoor temperature. For heating months, we model the heat supplied to the house as linear in $\Delta T^+$. Figure \ref{Bill_threshold} shows an example of the threshold $\tau^\star$ (dashed red vertical line) selected to minimize the sum of the RMSEs in the non-heating-month fit (green horizontal line below the threshold) and the heating-month fit (orange sloped line above the threshold).

\subsubsection{Uncertainty quantification} \label{BillMethodConfidenceInterval}

The analytical approach in Section \ref{Thermostat_design_CI} to constructing CIs works poorly for the bill method because smaller sample sizes mean fewer the degrees of freedom and larger critical $t$-values and CIs. For the one-year bill-method datasets, which typically contain no more than seven heating months, CIs can be very wide (the largest analytical CI in our dataset spanned over 23.5 kW [80,000 Btu/h]). To address this, we use the Bayesian bootstrap \cite{Rubin1981} to construct empirical CIs. The Bayesian bootstrap works by adding a layer of uncertainty in the form of $n$ randomly generated weights drawn from a $\text{Dirichlet}(1,\dots,1)$ distribution, where $n$ is the number of data points. At each iteration, we generate a new set of $n$ weights and perform a weighted least squares fit on the full dataset. We use the resulting model to calculate a point estimate of the design load. We repeat this procedure a number of times, then construct the CI by taking the $\alpha/2$ and $1 - \alpha/2$ quantiles of the set of point estimates, where $\alpha$ is the desired confidence level. The Bayesian bootstrap method helps for small sample sizes because the randomly generated weights come from a continuous distribution, creating a smoother posterior distribution and more stable tail behaviors than classical bootstrap methods that use discrete resampling.

\section{Results}\label{results}

\begin{figure*}
\centering
\includegraphics[width=0.69\textwidth]{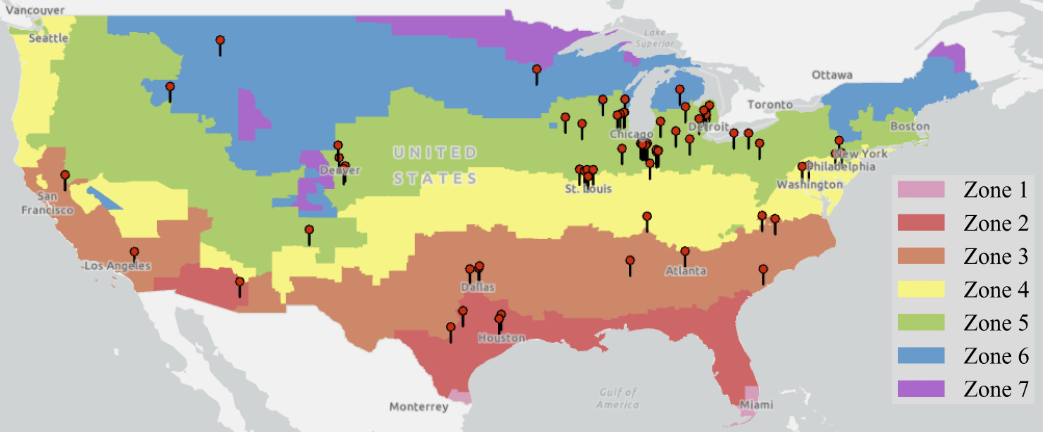}
\includegraphics[width=0.29\textwidth]{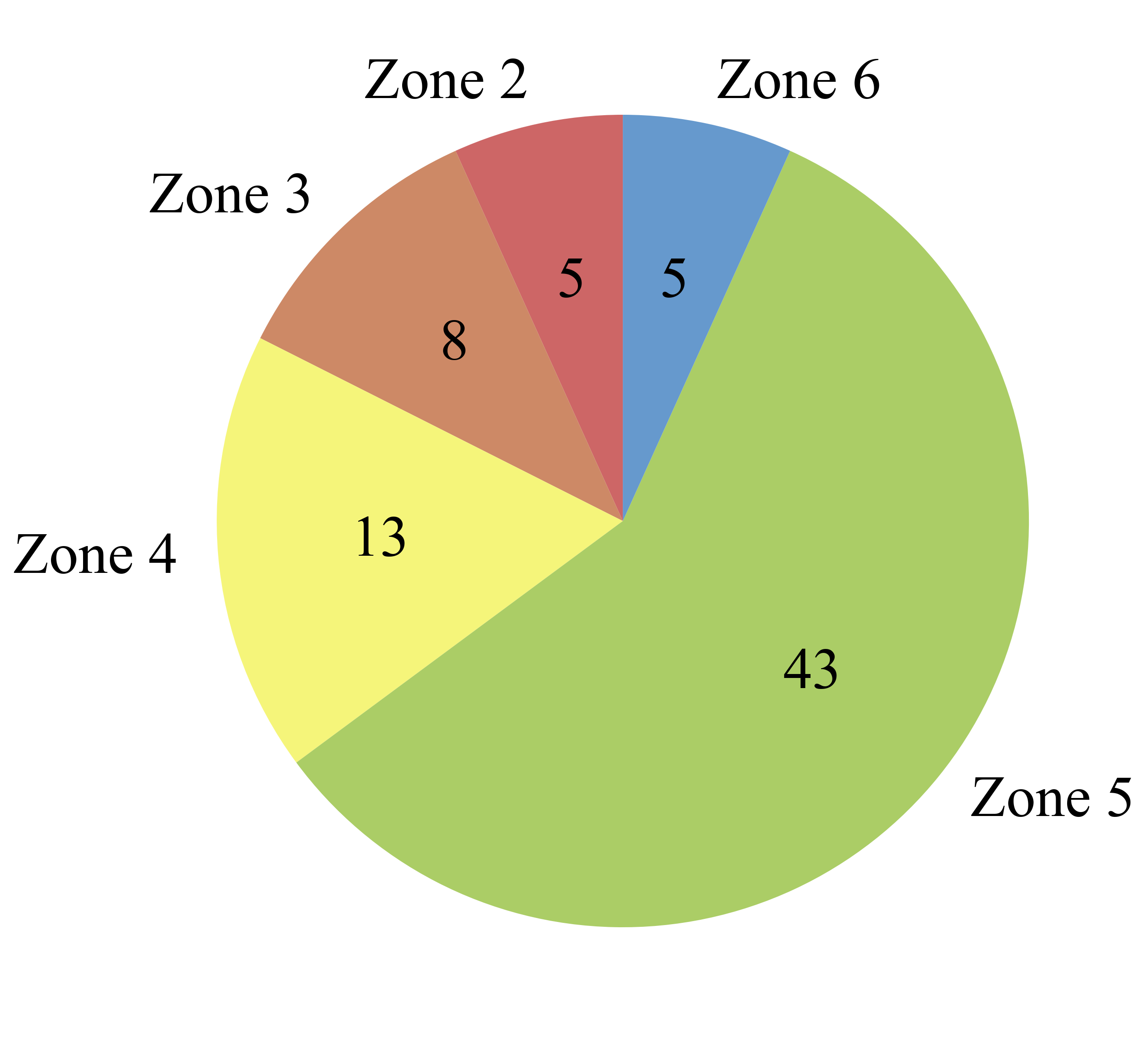}
\caption{The 74 participating houses span five United States climate zones. The majority are in zone five, a cold climate.}
\label{fig:locations}
\end{figure*}

This section summarizes the dataset demographics and compares several design load estimates from the thermostat method, the bill method, the Manual J calculations we purchased, a commercial software package that uses proprietary methods \cite{Rookstack}, and a rule of thumb that prescribes a climate-dependent amount of design heating load per unit of conditioned floor area.

\subsection{Survey demographics}

\begin{figure*}
\centering
\includegraphics[width=0.47\textwidth]{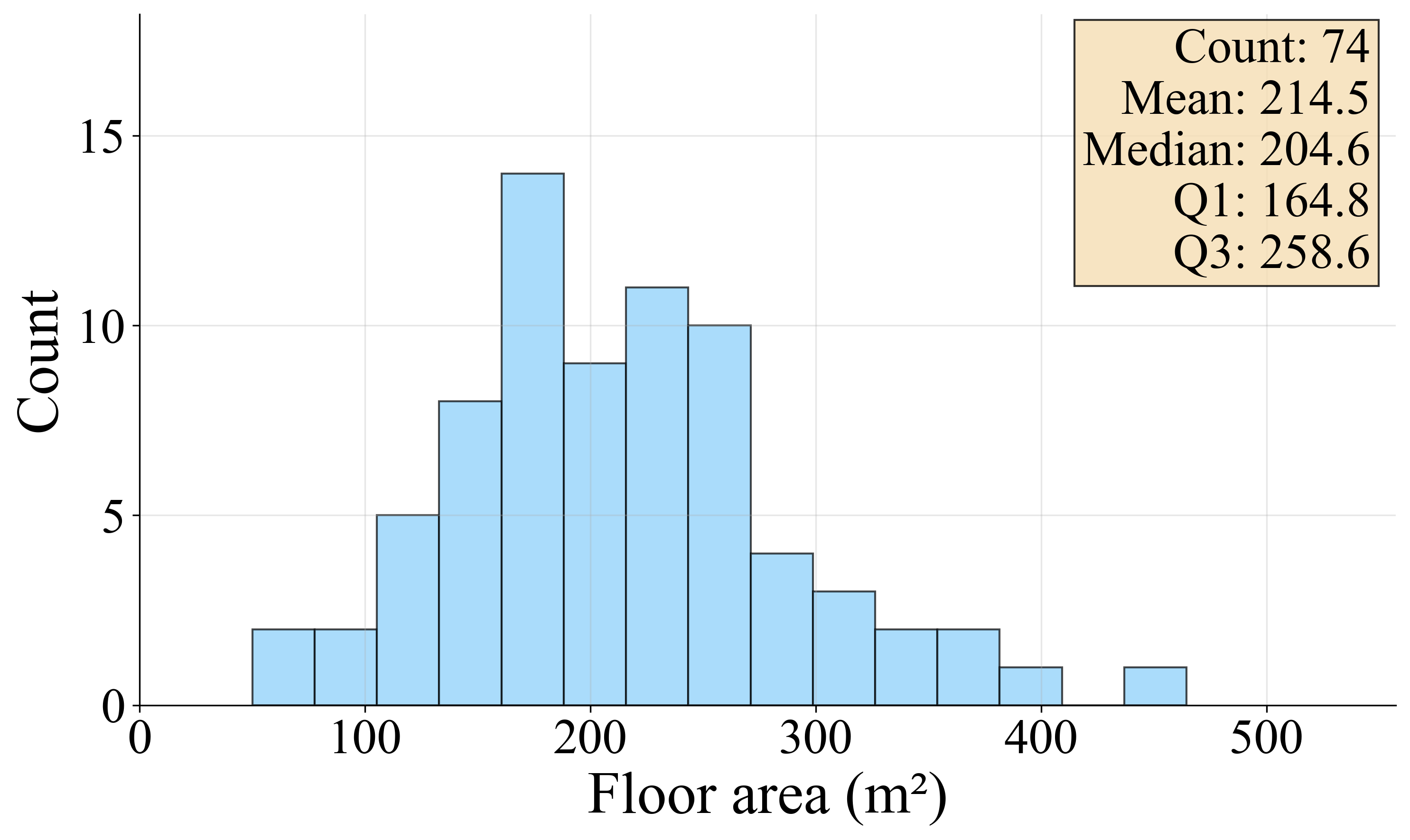} \qquad
\includegraphics[width=0.47\textwidth]{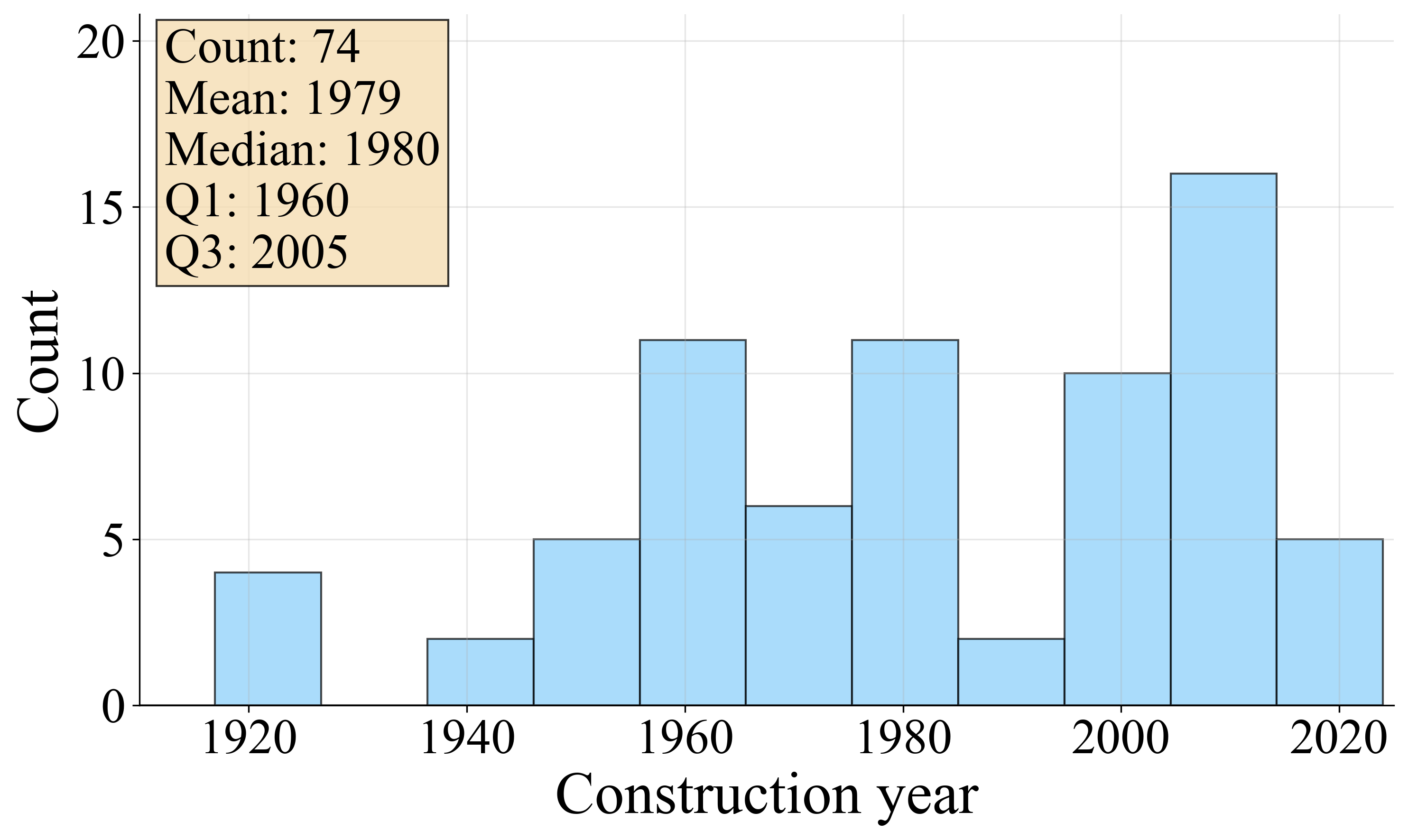}
\caption{Histograms of the floor area (left) and construction year (right) for the participating houses.}
\label{fig:floor_area_construction_year}
\end{figure*}

Figure \ref{fig:locations} shows participant locations and American Society of Heating, Refrigerating and Air-Conditioning Engineers climate zones. All participating houses are located in the United States. Participant locations span climate zones two through six and 29$^\circ$ to 47$^\circ$ in latitude. Of the 74 houses, 43 (58\%) are in zone five, a cold climate with 3,000 to 4,000 heating $^\circ$C-days per year at a balance-point temperature of 18 $^\circ$C. Figure \ref{fig:floor_area_construction_year} shows floor area and construction year histograms. The majority of houses have between 165 and 262 m$^2$ (1,776 and 2,820 ft$^2$) of floor area and were built between 1960 and 2005.

\subsection{Thermostat method results} \label{thermostat_results}

The left histogram in Figure \ref{fig:R2_histograms} shows the coefficient of determination ($R^2$) values for the thermostat method fit on each participant's data. The mean and median $R^2$ values are 0.89 and 0.90, respectively. This implies that, on average, about 90\% of the variation in the observed heat supply can be explained by the first-order linear impact of the indoor-outdoor temperature difference. The high $R^2$ values indicate good overall fit and support the use of first-order linear models for estimating design heating loads.

For \PercentOfInterpolating\% of participants, the thermostat data contained at least one night where the average indoor-outdoor temperature difference was as large as the design temperature difference. For these houses, the design heating load estimate was an interpolation of the dataset, rather than an extrapolation---the dataset essentially contained a measurement of the design load, up to possible inaccuracies in the heating capacity listed in the equipment specification sheet. We also performed 10-fold cross validation using the thermostat method for the 68 out of 74 participants with enough data (defined as 50 data points, leaving at least 5 validation points for each of the 10 folds). The RMSE and $R^2$ results showed good agreement between the training and validation datasets.

Time-averaging the thermostat data over each night, as outlined in Section \ref{thermostat_time_average}, significantly improves the model fit. With only hourly averaging, the mean and median $R^2$ values drop to 0.56 and 0.57, respectively. Averaging over entire days (midnight to midnight) only slightly improves the $R^2$ statistics compared to overnight averaging, an effect we attribute to smoother data averaged over longer time windows (24 hours vs. about eight) being easier to predict. However, averaging the thermostat data over entire days also roughly doubled the estimated exogenous heat gains. This is expected, as daytime data include solar heat gains. We chose to include only nighttime data in the model fits because peak heating loads typically occur a few hours before sunrise, when solar heat gains are zero and occupant activity is low.

\subsection{Bill method results} \label{bill_results}

\begin{figure*}
    \centering
    \includegraphics[width=0.47\textwidth]{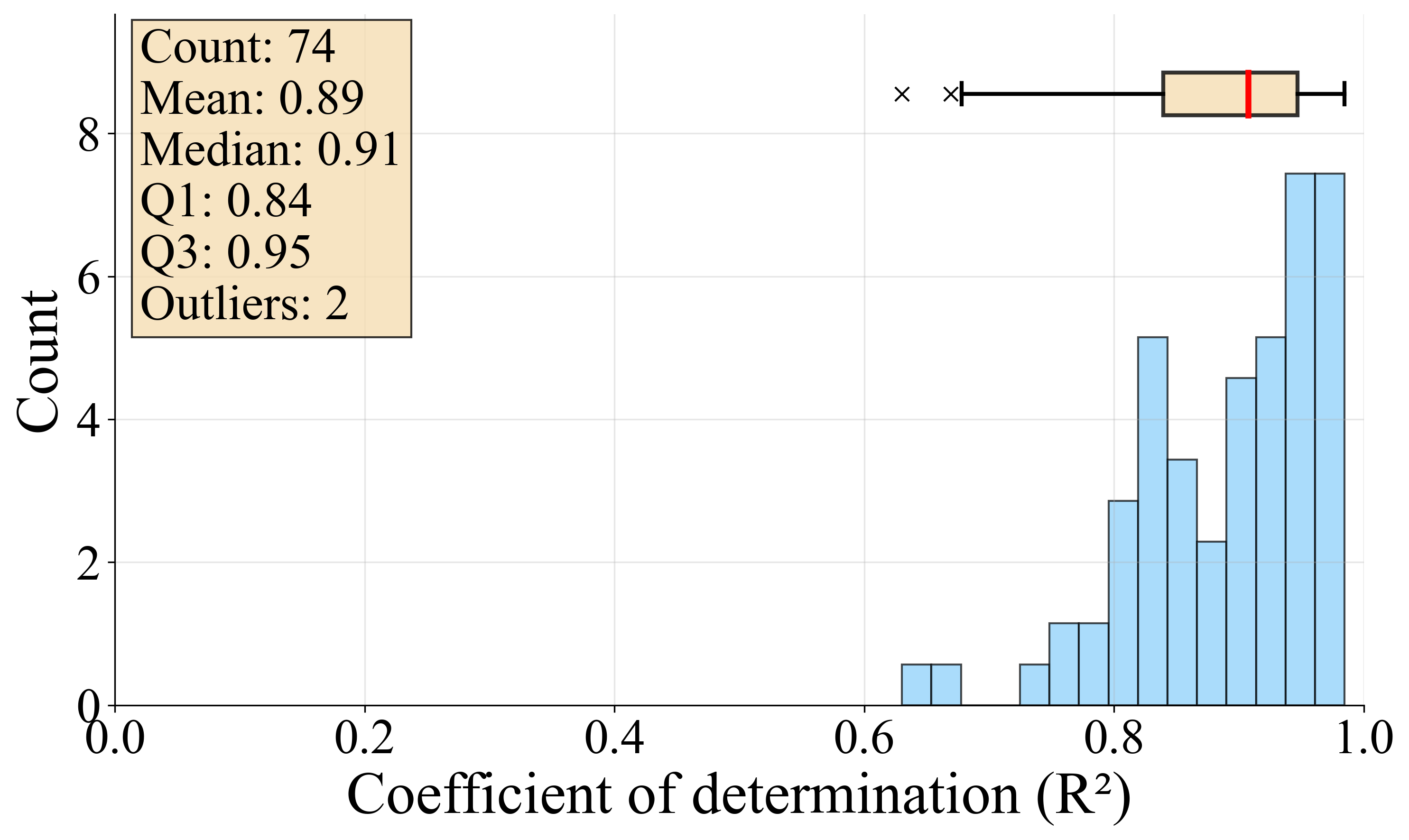} \qquad
    \includegraphics[width=0.47\textwidth]{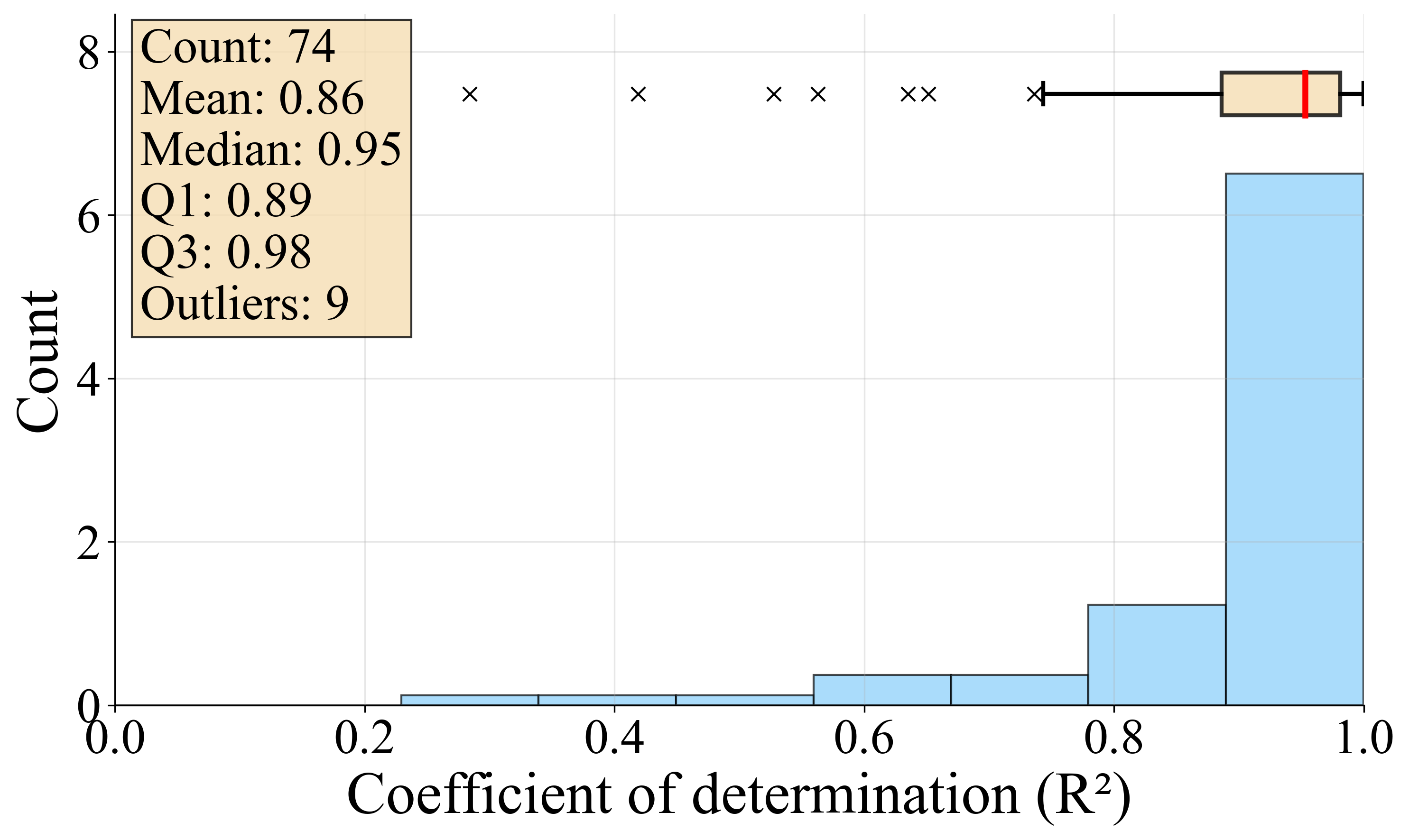}
    \caption{The thermostat method (left histogram) has a lower median $R^2$ than the bill method (right) but fewer outliers.}
    \label{fig:R2_histograms}
\end{figure*}

The righthand histogram in Figure \ref{fig:R2_histograms} shows the $R^2$ values for the utility bill method fits. The mean and median (0.90 and 0.95, respectively) are higher than for the thermostat method, but there are more outliers toward the lower end of the distribution. We attribute these outliers to the fact that the bill method uses significantly fewer data points. In addition, the bill data, which are averaged over months rather than single nights, also exhibit higher variability because heating patterns between the shoulder seasons and the coldest winter months vary significantly from participant to participant. For example, some participants may choose to keep the heating system turned off entirely during shoulder seasons to reduce energy bills or pollutant emissions.

The Bayesian bootstrap method discussed in Section \ref{BillMethodConfidenceInterval} significantly tightens the 95\% CIs relative to the analytical method based on the $t$-distribution. On average over the fits for all participants, the Bayesian bootstrap decreases the CI width by about 50\%. Narrower CIs should enable more accurate sizing of heating equipment.

\subsection{Method comparisons} \label{results_meta_analysis}

To evaluate how closely the thermostat and bill methods agree, we performed two statistical hypothesis tests. The strictest test evaluates whether the central design heating load estimates from both methods are equal. At the 95\% confidence level, the thermostat and bill methods met this strict criterion for 36\% of houses. We also conducted a more lenient test of whether the point estimates from the two methods differ by no more than $\pm\delta$ (kW). With tolerances $\delta = 1.76$ and 3.52 kW (6,000 and 12,000 Btu/h---typical capacity increments in which residential HVAC equipment is manufactured), the thermostat and bill methods met this criterion at the 95\% confidence level for 71\% and 80\% of the houses, respectively.

\begin{figure}
    \centering
    \includegraphics[width=0.5\textwidth]{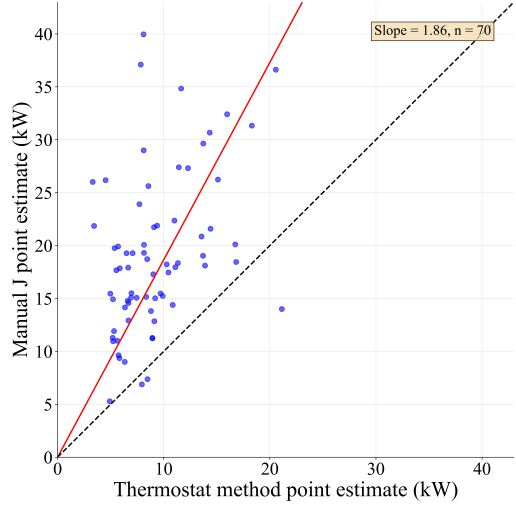}
    \caption{Parity plot comparing design heating load estimates for the Manual J and thermostat methods. Manual J estimates are roughly two times higher on average than the thermostat and bill method estimates, respectively.}
    \label{fig:thermostat_bill_comparison}
\end{figure}

The parity plot in Figure \ref{fig:thermostat_bill_comparison} compares point estimates of design heating loads from Manual J and the thermostat method. The bill method estimates are not pictured, but generally agree well with the thermostat method. On average, the Manual J estimates are \ManualJToThermostat\ and \ManualJToBill\ times higher than the corresponding estimates from the thermostat and bill methods, respectively.

Table \ref{tab:results_percent_diff_thermostat_bill} show the ratios (means and 95\% CIs over the 74 houses) of design heating load estimates from existing methods to the thermostat- and bill-method estimates. All existing methods produce estimates that are significantly higher than the data-driven estimates. The commercial software estimates are consistently the lowest of the existing methods, averaging \RookStackToThermostat\ and \RookStackToBill\ times higher than the thermostat-method and bill-method estimates, respectively. On average, the Manual J estimates are more than double the estimates from either data-driven method. Manual J produces modestly lower estimates than like-for-like replacement. Existing equipment is on average \ExistingCapToThermostat\ and \ExistingCapToBill\ times larger thermostat-method and bill-method estimates, respectively. We based the rule-of-thumb estimates on floor area and climate zone, using the W/m$^2$ values recommended by several furnace sizing websites. The values ranged from 103 W/m$^2$ (32.5 Btu/h per ft$^2$) for zone one to 189 W/m$^2$ (60 Btu/h per ft$^2$) for zone seven. This rule of thumb produces the highest estimates by far.

\section{Discussion}
\label{discussion}

\subsection{Lessons learned from data collection}
\label{discussion_dataCollection}

The dataset collected in this study enabled the development and comparison of methods for design load estimation. This section summarizes five lessons learned from data collection. First, the amount of work required to interact with the participants should not be taken lightly. From our initial outreach to roughly 5,000 people, only 74 participants ultimately provided all of the requested data. We estimate that we exchanged over 1,000 emails with individual participants, not counting outreach through large mailing lists. Data collection took over six months.

Second, we found it effective to structure data requests sequentially and provide gift cards to incrementally reward completion of each step. We began with a housing characteristics survey to enable screening for anomalies such as unusual heating equipment or untracked secondary heating sources like space heaters. Each participant received a \$10 gift card for completing the survey. Next, we asked participants to share smart thermostat data (or remote thermostat access), utility bills, and a photo of their heating equipment nameplate. Each participant received a \$20 gift card for submitting all three. We commissioned Manual J calculations at the end of the data collection process, since this was the most expensive and time-consuming task. Leaving Manual J until the end also lowered the risk of participants dropping out of the process after receiving their Manual J calculations. 

Third, finding qualified individuals to perform Manual J calculations was a real challenge. We reached out to over 100 potential providers in 18 US cities, and less than 20\% provided quotes for Manual J services. The quotes we received varied from \$300 to \$2,500. 

Fourth, the most challenging data to collect were the heating equipment nameplate photos, which enabled us to look up equipment efficiencies and heating capacities. Participants, who had no particular HVAC expertise, often understandably confused the central air conditioner or air handler for the furnace. In many cases, multiple email exchanges were required to identify the correct nameplate. This part of the study was also where we lost the most participant engagement, likely because heating equipment often occupies inconvenient spaces. 

Fifth, we found it effective to connect with existing communities that share interests in home energy. For example, we found an online community that helps users visualize historical data from a popular brand of smart thermostat \cite{beestat}. Community members already owned smart thermostats and were interested in home energy monitoring, which significantly boosted their engagement in our study. About 75\% of the participants that completed our study heard about it through this online community.

\begin{table*}
    \centering
    \caption{Ratios of the design heating load estimates from existing methods to the thermostat- and bill-method estimates. Means and CIs are taken over the 74 houses in the dataset.}
    \label{tab:results_percent_diff_thermostat_bill}
    \small
    \setlength{\tabcolsep}{4pt}
    \begin{tabular}{l | c c | c c}
        & \shortstack{ \textbf{Ratio mean} \\ \textbf{(thermostat)}} & \shortstack{ \textbf{Ratio 95\% CI} \\ \textbf{(thermostat)} } & \shortstack{ \textbf{Ratio mean} \\ \textbf{(bill)}} & \shortstack{ \textbf{Ratio 95\%} \\ \textbf{CI (bill)} } \\
        \hline
        \textbf{Commercial software} & \RookStackToThermostat & [\RookStackToThermostatLow, \RookStackToThermostatHigh] & \RookStackToBill & [\RookStackToBillLow, \RookStackToBillHigh] \\
        \textbf{Manual J} & \ManualJToThermostat & [\ManualJToThermostatLow, \ManualJToThermostatHigh] & \ManualJToBill & [\ManualJToBillLow, \ManualJToBillHigh] \\
        \textbf{Like-for-like replacement} & \ExistingCapToThermostat & [\ExistingCapToThermostatLow, \ExistingCapToThermostatHigh] & \ExistingCapToBill & [\ExistingCapToBillLow, \ExistingCapToBillHigh] \\
        \textbf{Floor-area rule of thumb} & \HeuristicToThermostat & [\HeuristicToThermostatLow, \HeuristicToThermostatHigh] & \HeuristicToBill & [\HeuristicToBillLow, \HeuristicToBillHigh] \\
\end{tabular}
\end{table*}

\subsection{Implementation practicalities}

We view the thermostat method as more robust than the bill method, as the thermostat method uses a larger dataset with higher temporal resolution (sub-hourly vs. monthly). We also found that thermostat data contained observations from design conditions for about three-quarters of participants, making the thermostat method estimate nearly a measurement of the design heating load (up to possible inaccuracies in heating equipment capacity specifications).

However, the thermostat method requires a smart thermostat that logs historical data and allows data export. Such a thermostat typically costs around \$140 \cite{ecobee_thermostat}. The thermostat method can also be challenging to implement for households with highly dynamic temperature setpoint patterns. Some households keep a steady setpoint, some have a routine nighttime setbacks, and others keep highly irregular setpoint patterns. While selection of appropriate averaging windows for the thermostat method could in principle be automated, we found it less time-intensive simply to plot the setpoint time series and manually select appropriate windows. Finally, the thermostat method requires an estimate of the existing equipment's heating capacity. Most participants in our study did not know this information, so we obtained it indirectly by asking for a photo of the equipment nameplate, from which we could read off or look up the heating capacity. Obtaining nameplate photos was among the most challenging aspects of data collection, as discussed in Section \ref{discussion_dataCollection}.

The bill method is significantly simpler to implement than the thermostat method. The bill method only requires monthly energy bills, which most households have or can access through utility websites. The bill method also requires only the efficiency of the existing equipment, not its heating capacity. While obtaining the nameplate efficiency presents the same challenges as obtaining the nameplate capacity, most households can provide a qualitative assessment of their heating equipment's age and quality (e.g., old or new, low- or high-end), from which the efficiency can be estimated.

However, the bill method has limitations related to small sample sizes. Given a year of utility bills, the bill method typically uses only six or seven data points, each corresponding to heating fuel use over a month with nontrivial heat demand. Small sample sizes can make estimation results volatile and sensitive to outliers. The Bayesian bootstrap method used in this paper mitigates the bill method's robustness issues but does not eliminate them. Some households also forgo heating in late fall or early spring, or when away on vacation, which affects monthly fuel use patterns. Unlike the thermostat method, which in many cases interpolates heating data to observed design conditions, the bill method requires significant extrapolation since monthly average outdoor temperatures are typically much less extreme than design conditions. Despite these shortcomings, we found that the bill method ultimately provides similar design load estimates to the thermostat method (see Section \ref{results_meta_analysis} and Figure \ref{fig:thermostat_bill_comparison}). For these reasons, we view the bill method as better suited to low-cost implementation at scale.

The commercial software package that we tested provides estimates that are generally midway between Manual J and the thermostat and bill methods (on average, \RookStackToThermostat \ and \RookStackToBill \ times higher than the thermostat and the bill methods, respectively). The commercial software also does not require a site visit, only a remote housing characteristics survey that takes 10 to 20 minutes to complete.

\subsection{Limitations and future work}
\label{discussion_limitations}

One limitation of this study is the structure of the dataset we gathered. The participants who completed all of our data-sharing requests tended to be home energy enthusiasts. For example, nearly three out of four heard about our study through an online community centered on visualizing smart thermostat data. This possible source of selection bias could mean that our participants have invested more in energy efficiency retrofits (such as insulation and air-sealing) than most households, possibly making their design loads unusually low for their floor areas and climate zones. This effect could partly explain the very high ratios of the design-load estimates from the floor-area rule of thumb to the data-driven methods. It could also partly explain the very high ratios of existing equipment capacities to the data-driven design load estimates. 
We note, however, that both Manual J and the commercial software package we tested should in principle reflect the house as built, including any energy efficiency retrofits.

Although we did not intentionally select for detached single-family houses, we did not receive full datasets for any attached or multi-family housing. While we designed the methods in this paper to work with all housing types, testing the methods on data from attached housing, such as apartments and row houses, is an important direction for future work. All of the houses in our dataset were in the United States---disproportionately in climate zone five, a cold climate---and heat with gas furnaces. Future work could test data-driven design load estimation methods in other climates or with other equipment, or compare data-driven methods to industry standards from other countries.

Another limitation is the lack of ground-truth measurements of the actual design loads. While the thermostat data for \PercentOfInterpolating\% of the houses in this study contained observations from heating design conditions, making those thermostat-method estimates nearly measurements of the design loads, new methods for estimating design loads should in principle be benchmarked against ground-truth measurements. In practice, however, accurately measuring heat transfer rates from HVAC equipment requires flow and temperature sensors that can be costly and invasive to install; for cooling equipment, humidity sensing presents additional challenges. One possible path around thermal instrumentation challenges is to install circuit-level electricity metering---which is generally less costly, less invasive, and more accurate---in housing with electric resistance heating, for which the thermal power output in steady state equals the electrical power input.

This study could be extended in at least four other ways. First, this study's scope included only heating design loads. Data-driven estimation of design cooling loads is an important but more challenging direction for future work, as it involves not only heat transfer through the building envelope, but also solar heat gains, humidity effects, and temperature-dependent equipment COPs and capacities. Second, this study considered only design load estimation. Future work could investigate data-driven methods for mapping design heating and cooling load estimates or CIs into equipment sizing recommendations. Third, the utility bill method developed here used only natural gas bills, but it could be extended to accommodate housing with electric resistance heating or heat pumps. Fourth, researchers could work with standard-setting agencies to incorporate data-driven methods for design load estimation into industry guidelines for HVAC equipment sizing.

\section{Conclusion}
\label{conclusion}

This paper developed two data-driven methods for estimating design heating loads and tested the methods on the richest dataset of which the authors are aware. This paper found that both data-driven methods---one based on thermostat data, the other on monthly energy bills---had strong goodness-of-fit statistics individually and agreed reasonably well with each other. Relative to the data-driven methods investigated here, the Manual J standard, as implemented in the real world by practicing professionals, overestimated design heating loads by more than a factor of two. Other existing methods, such as rules of thumb based on floor area or like-for-like replacement, produced even higher design-load estimates.

\section*{Acknowledgements}

This material is based upon work supported by the United States Department of Energy, Office of Science, Building Technologies Office, under Award Number DE-SC0025089. The authors also acknowledge support from the Applied Research Institute (ARI) and the Indiana Economic Development Corporation (IEDC). The authors thank all the participants who provided the data for this study and contacts at the Northwest Energy Efficiency Alliance, BeeStat, and Rewiring America for helpful discussion. We thank Mark Ladd, founder and CEO of Convectiv, for providing design load estimates from the commercial software cited in this paper.

\bibliographystyle{elsarticle-num}
\bibliography{biblio.bib}

\end{document}